\documentclass[fleqn,twoside]{article}
\usepackage{gc,cite}

\heads
{Saibal Ray and Basanti Das}
{Relativistic gravitational mass in Tolman-VI solution}

\begin{document}
\twocolumn[
\Arthead{13}{2007}{3 (51)}{224}{230}

\bigskip

\Title{Relativistic gravitational mass in Tolman-VI solution}

\Authors{Saibal Ray\foom 1}
        {and Basanti Das\foom 2}
{Department of Physics, Barasat Government College, Kolkata
	700 124, North 24 Parganas, West Bengal, India;
	Inter-University Centre for Astronomy and Astrophysics, Post Bag
	4, Ganeshkhind, Pune 411 007, India}      
{Belda Prabhati Balika Vidyapith, Belda, Midnapur 721 424, West Bengal,
 India}     

\Rec{23 May 2006}
\Recfin{20 June 2007}

\Abstract
    {Some known solutions for static charged fluid spheres of Tolman-VI
 type
    in general relativity are reexamined. The gravitational mass that
    appears in the Pant and Sah \cite{pan79} and Tikekar \cite{tik84}
    solutions is shown to be of electromagnetic origin in the sense
 that the
    gravitational mass, along with all other physical quantities,
 depends on
    the electromagnetic field alone. The energy condition, singularity
 and
    stability problems of the models are discussed thoroughly.}

\RAbstract
    {¥«ïâ¨¢¨áâáª ï £à ¢¨â æ¨®­­ ï ¬ áá  ¢ VI à¥è¥­¨¨ ’®«¬¥­ }
    {‘ ©¡ «  ©,  § ­â¨ „ á}
    {¥ª®â®àë¥ ¨§¢¥áâ­ë¥ à¥è¥­¨ï Ž’Ž â¨¯  VI ’®«¬¥­  ¤«ï § àï¦¥­­ëå
 ¦¨¤ª¨å
    áä¥à  ­ «¨§¨àãîâáï § ­®¢®. ®ª § ­®, çâ® £à ¢¨â æ¨®­­ ï ¬ áá  ¢
 à¥è¥­¨ïå
     ­â  ¨ ‘  [1] ¨ ’¨ª¥ª à  [2] ¨¬¥¥â í«¥ªâà®¬ £­¨â­®¥ ¯à®¨áå®¦¤¥­¨¥
 ¢ â®¬
    á¬ëá«¥, çâ® á ¬  £à ¢¨â æ¨®­­ ï ¬ áá , ª ª ¨ ¤àã£¨¥ ä¨§¨ç¥áª¨¥
 ¢¥«¨ç¨­ë,
    § ¢¨áïâ â®«ìª® ®â í«¥ªâà¨ç¥áª®£® § àï¤ . ®¤à®¡­® ®¡áã¦¤ îâáï
    í­¥à£¥â¨ç¥áª¨¥ ãá«®¢¨ï ¨ ¯à®¡«¥¬ë á¨­£ã«ïà­®áâ¨ ¨ ãáâ®©ç¨¢®áâ¨
 ¬®¤¥«¥©.}

] 
\email 1 {saibal@iucaa.ernet.in}

\section{Introduction}

  Studies of static solutions in connection with stellar interiors have
 been
  of continuing interest for many researchers in the framework of
 general
  relativity. These works can be divided into two categories: (1)
 static
  fluid spheres without charge and (2) charged static fluid spheres.
 Some
  remarkable works on the neutral cases of Einstein's field equations,
  related to the Schwarzschild solution, were done by Tolman
 \cite{tol39},
  Wyman \cite{wym49}, Leibovitz \cite{lei69}, Whitman \cite{whi77} and
 Bayin
  \cite{bay78}. On the other hand, solutions of the coupled
 Einstein-Maxwell
  field equations have been obtained by several authors. Some of the
  interesting works in this direction with the Reissner-Nordst{\"o}m
  space-time can be found out in the papers by Weyl \cite{wey17},
 Majumdar
  \cite{maj47}, Papapetrou \cite{pap47}, Bonnor \cite{bon65}, Kyle and
  Martin \cite{kyl67} and Cooperstock and de la Cruz \cite{coo77}.

  The reason behind the work on the coupled charge-matter distributions
 is
  mostly to avert the singularities that occur in the
 Schwarzschild-like
  solutions. These singularities are of two types: (1) the coordinate
  singularity at $r=2m$ which provides the event horizon of a black
 hole
  of mass $m$, and (2) the physical singularity at $r=0$, the centre of
 the
  spherical system. Now, in connection with the singularity, it is
 observed
  that the gravitational collapse of a spherically symmetric
 distribution of
  matter to a point singularity may be avoided in the presence of
 charge.
  This happens since the repulsive Coulomb force due to the charge (in
  addition to the thermal pressure gradient in the fluid) opposes the
  gravitational attraction. Therefore, the issue of studying the
  Einstein-Maxwell space-time has always received considerable
 attention
   from the researchers \cite{pan79,fel95,sha01,iva02}.

  However, there was another issue in studying charged static fluid
 spheres,
  which is associated with {\it electromagnetic mass models}. According
 to
  Lorentz \cite{lor04}, in an extended electron ``there is no other, no
  `true' or 'material' mass'', and this provides only an
 ``electromagnetic
  masses of the electron''.  Wheeler \cite{whe62} also believed that
 the
  electron has a `mass without mass', while Feynman \cite{fey64} termed
 this
  type of models the ``electromagnetic mass models'', where mass
 originates
  from the electromagnetic field alone and hence gives a
 phenomenological
  relationship between the gravitational and electromagnetic fields
  \cite{flo77,coo77,tiw84,gau85,gri86,leo87}.

  Therefore, motivated by the above ideas, we have recently
 \cite{ray04a}
  investigated the class of astrophysical solutions previously obtained
 by
  us \cite{ray02} and also by Pant and Sah \cite{pan79} for a static,
  spherically symmetric Einstein-Maxwell space-time. While discussing
 the
  Pant and Sah solutions \cite{pan79}, which represent the charged
 analogue
  of the Tolman VI solution \cite{tol39}, we \cite{ray04a} have studied
 the
  case $n = 0$ only related to the electromagnetic mass models
  \cite{lor04,fey64}, where $n$ is a free parameter appearing in the
 set of
  solutions. In the present investigations, however, it is shown that
 the
  electromagnetic mass models can also be constructed not only for the
  specific choice $n = 0$ but for the whole range of $0 \leq n \leq 2$,
 even
  including fractional values.

  It is known that the Tikekar solutions \cite{tik84} represent a
  generalization of the Pant and Sah solutions \cite{pan79}, so we have
  also studied them as a continuation of the verification scheme. It
  is possible to show, from the expression for the charge and the
 equation
  of state, that the gravitational mass depends on the electric charge
 alone.

  In the concluding part, it is discussed that there are some other
  solutions available in the literature where, especially in the case
  ($\lambda, n $) in section X of Ivanov's paper \cite{iva02}, our
  requirement for a solution to describe an electromagnetic mass model,
 is
  also satisfied. For physical validity related to all cases studied in
 this
  article, the singularity problem, stability situation and the energy
  conditions are discussed in connection with de Sitter space-time.

\section{The Einstein-Maxwell field equations}

  Consider the spherically symmetric line element
\bear
	ds^{2}= g_{ij}dx^{i}dx^{j} = e^{\nu(r)} dt^{2} - e^{\lambda(r)}
	dr^{2} - r^{2} d \Omega^2,
\ear
  with $d\Omega^2 = d \theta ^{2} + \sin^{2} \theta\, d\phi^{2}$ in the
  standard coordinates $x^{(0, 1, 2, 3)} = (t, r, \theta, \phi)$, where
  $\nu$ and $\lambda$ are two functions of the radial coordinate $r$.

  The Einstein-Maxwell field equations corresponding to spherically
  symmetric static charged source are then given by
\bearr
	\e^{-\lambda} (\lambda'/r - 1/r^{2} ) + 1/r^{2} = 8\pi \rho + E^{2},
\yyy
	\e^{-\lambda} ( \nu'/r + 1/r^{2} ) - 1/r^{2} = 8\pi p -E^{2},
\yyy
	\e^{-\lambda} [{\nu''}/2 + {\nu'}^2/4 -
		{\nu' \lambda'}/4 + (\nu' - \lambda' )/ 2r]
\nnn \inch
	= 8\pi p + E^{2},
\yyy
	{(r^2 E)}' = 4\pi r^2 \sigma \e^{\lambda/2},
\ear
  where $\rho$, $p$, $E$ and $\sigma$ are the matter energy density,
 the
  fluid pressure, the electric field intensity and the electric charge
  density, respectively. Here, the prime denotes a derivative with
 respect
  to the the radial coordinate $r$.

  \eq (5) can be equivalently, in terms of the electric charge $q$,
  expressed as
\bear
	q(r) = r^2E(r) = \int_{0}^{r} 4 \pi r^2 \sigma \e^{\lambda/2} dr.
\ear
  Therefore, \eq (2), by the use of the above equation (6), reduces to
\bear
	\e^{-\lambda} = 1 - 2M(r)/r,
\ear
  where $M(r)$ is the active gravitational mass, which can be expressed
 in
  terms of the effective gravitational mass, $m(r)$, in the form
\bear
	m(r) = M(r) + \mu(r) = 4 \pi \int_{0}^{r}[\rho + {q^2}/{8 \pi r^4}]
	r^2 dr,
\ear
  $\mu(r)$ being the {\it mass equivalent of the electric field}. It is
 to
  be mentioned here that customarily people consider $\mu(r)$ as the
 {\it
  mass equivalent of electromagnetic field\/} because of the fact that
 it is
  associated with the idea of Lorentz's conjecture of `Electromagnetic
 mass'
  for the extended electron \cite{lor04,fey64}. However, the models
  considered here being static, the fields are not electromagnetic but
 only
  electric.

  It is interesting to note that \eq (8) provides the increase of the
 total
  gravitational mass due to inclusion of the charge
 \cite{coo77,flo77,her85},
  and in the absence of an electric charge it reduces to the usual
 active
  gravitational mass. In this regard, it is also to be noted that \eq
 (8) is
  in the form of Bekenstein's \cite{bek71} \eq (31), where the part of
 mass
  related to the fluid is irreducible and charge-independent.
 Therefore, in
  the next section, one of our aims is to find out the criteria for
  constructing electromagnetic mass models by showing that the matter
 part
  depends on the charge, and thus the entire gravitational mass turns
 out to
  be completely electromagnetic by origin.

\section{Gravitational masses of purely electromagnetic origin}

\subsection {The Pant-Sah models}

  For a static, spherically symmetric distribution of charged fluid,
 Pant and
  Sah \cite{pan79} obtained a class of solutions which are as follows:
\bear
	\e^{\nu} \eql b r^{2n},
\yy
	\e^{-\lambda} \eql c,
\\
	\rho \eql \frac{1}{16\pi r^2}[1 - c(n - 1 )^2],
\\
	p \eql \frac{1}{16\pi r^2}[c(n+1)^2 -1],
\\
	\sigma \eql \pm  \frac{1}{4\pi r^2}
	 \left[\frac{c}{2}\{1 - c (1 + 2n - n^2)\}\right]^{1/2},
\\
	 E^2 \eql  \frac{1}{2 r^2}[1 - c(1 + 2n - n^2)],
\ear
  where
\bear
	b \eql {a}^{-2n}\left[1 - \frac{2m}{a} + \frac{{q}^2}{{a}^2}\right],
\\
	c \eql \left[1 - \frac{2m}{a} + \frac{{q}^2}{{a}^2}\right]
	= \left[1 - \frac{{2q}^2}{{a}^2}\right]( 1 + 2n - n^2)^{-1}.
\nnn
\ear
  With this expression of $c$, the above class of solutions, for the
  cosmological constant $\Lambda=0$ and also for the integration
 constant
  $B=0$, represents a charged analogue of the Tolman VI \cite{tol39}
  solution. It is also to be noted that the matter density, the fluid
  pressure, electric charge density and the radial electric field
 involved
  in \eqs (11)--(14) follow the inverse square laws.  Consequently,
 these
  physical parameters increase as the distance decreases and become
 infinite
  at the centre of the spherical system. Therefore, due to
 singularities at
  the centre, such models are not physical as possible descriptions of
  stellar structure.  They may be either regarded as the
  Reissner-Nordst{\"o}m black hole solutions, or one may think of a
  technique by which the singularity can be averted. We shall consider
 this
  point later.

  Now, the total gravitational mass $m(r=a)$ can be calculated from \eq
 (8)
  using \eqs (11) and (14):
\bear
	m = \frac{n(2 - n)a^2 + 2q^2}{2(1 + 2n - n^2)a}.
\ear
  A close observation of \eq (17) shows that the status of the
 gravitational
  mass depends on the parameters $n$, $a$ and $q$.  Therefore, once we
 fix
  the values of $a$ and $q$ for a given charged spherical system, $n$
 can be
  regarded as an adjustable parameter `having real, not necessarily
 integral,
  values' \cite{tol39}. However, for physical viability, the values to
 be
  assigned to this parameter are $0 \leq n \leq 2$. One can obviously
 assign
  innumerable values to $n$ within this range. Related to the case $n =
 0$,
  it has been already shown by us \cite{ray04a} that for vanishing
 electric
  charge all the physical quantities, including the gravitational mass,
  vanish, and the space-time becomes flat. We therefore, for our
 purpose,
  would like to study the following four subcases A--D only.

\subsubsection*{{\bf A. }$n = 0.5$}

  This case is favourable for some historical reasons. This choice was
  originally made by Tolman \cite{tol39} himself in his uncharged
 version.
  In this case, the gravitational mass becomes
\bear
	m = \frac{3a^2 + 8q^2}{14a}.
\ear
  The functional condition containing the energy density $\rho$ and the
  pressure $p$, by virtue of \eqs (11) and (12), can be written as
\bear
  	\rho + p = \frac{1}{4\pi r^2}\left[\frac{n(a^2 - 2q^2)}{(1 + 2n -
		n^2)a^2}\right],
\ear
  which for the present case reduces to
\bear
	\rho + p =\frac{1}{14\pi r^2}\left[1 - \frac{2q^2}{a^2}\right].
\ear

\subsubsection*{{\bf B. }$n = 1$}

  For this choice, the gravitational mass becomes
\bear
	m = \frac{a^2 + 2q^2}{4a},
\ear
  whereas the functional condition containing the energy density $\rho$
 and
  the pressure $p$ is
\bear
	\rho + p = \frac{1}{8\pi r^2}\left[1 - \frac{2q^2}{a^2}\right].
\ear
  It has been pointed out by Herrera and Ponce de Le{\'o}n \cite{her85}
 that
  this particular solution, $n = 1$, admits a one-parameter group of
  conformal motions and is homothetic.

\subsubsection*{{\bf C. }$n =1.5$}

  In this case, the gravitational mass is
\bear
	m = \frac{3a^2 + 8q^2}{14a},
\ear
  and the related functional condition here becomes
\bear
	\rho + p = \frac{3}{14\pi r^2}\left[1 - \frac{2q^2}{a^2}\right].
\ear
  The gravitational mass in this case is the same as for $n = 0.5$.

\subsubsection*{{\bf D. }$n = 2$}

  The gravitational mass in this case becomes
\bear
	m = \frac{q^2}{a}.
\ear
  This expression is precisely the same as that in the $n = 0$ case of
 our
  previous work \cite{ray04a}, it vanishes for vanishing electric
 charge and
  thus provides an `electromagnetic mass' model \cite{lor04,fey64}.
 However,
  the functional condition for the present situation differs from that
 of
  the $n = 0$ case and is given by
\bear
	 \rho + p = \frac{1}{2\pi r^2}\left[1 - \frac{2q^2}{a^2}\right].
\ear

  Let us now, following Ray and Das \cite{ray02, ray04a}, and Ivanov
  \cite{iva02}, choose a relation between the electric charge and the
 radial
  coordinate of the fluid distribution as follows:
\bear
	q(r) = Kr^s,
\ear
  where $K$ and $s$ are two free parameters. For the specific choice
 $K=
  1/\sqrt 2$ and $s = 1$, the {\it ansatz\/} expressed in \eq (27)
 reduces
  to $q(a)/a = 1/\sqrt{2}$, where $a$ is the radius of the charged
 sphere.
  It is interesting to note that for this charge-radius ratio all the
 above
  perfect fluid functional conditions reduce to the form $\rho + p =
 0$.
  This is known as the `pure charge condition' \cite{gau85} and also
 the
  imperfect-fluid equation of state in the literature since the matter
  distribution under consideration is in tension and hence the matter
 is
  named as a `false vacuum' or `degenerate vacuum' or `$\rho$-vacuum'
  \cite{dav84,blo84,hog84,kai84}. We would like to comment here that
  this imperfect-fluid equation of state precisely describes the
  cosmological constant, viz., the vacuum energy, and by the phrase
 `pure
  charge condition' Gautreau [22] indicates the necessary condition for
  the `Lorentz-type pure-charge extended electron' to be of
 electromagnetic
  origin.

  The above charge-radius ratio in turn makes all the total effective
  gravitational mass vanish for vanishing charge. Therefore, the
  gravitational masses here are of purely electromagnetic origin. It is
 to
  be noted here that all the above cases $n = 0.5, 1, 1.5$, including
 our
  previous case $n = 0$ \cite{ray04a}, provide `electromagnetic mass'
 models
  with the imperfect-fluid equation of state. The only exception here
 is the
  last case $n = 2$ where we have obtained an electromagnetic mass
 model
  even under the perfect fluid condition. Although for the
 charge-radius
  ratio $q(a)/a = 1/\sqrt{2}$, the perfect fluid functional condition
 in \eq
  (26) reduces to the imperfect one, but this {\it ansatz\/} is not
  necessary for making the gravitational mass in \eq (25) suitably
 vanish as
  in the other cases.

  Let us look at \eq (26) differently. The components of this equation,
  viz., the matter energy density and the fluid pressure are given by
\bear
	\rho \eql \frac{1}{8\pi r^2}\frac{q^2}{a^2},
\\
	p \eql \frac{1}{8\pi r^2}\left[4 - \frac{9q^2}{a^2}\right].
\ear
  It can be checked that for a positive pressure the constraint on the
  charge-radius ratio to be imposed here is $q(a)/a \leq \pm 2/3$. The
  pressure vanishes for the value $\pm 2/3$, whereas it becomes
 positive for
  values less than $\pm 2/3$. Thus we have come across a case for
  electromagnetic mass models where the equation of state is a
 perfect-fluid
  one with a positive pressure. This case is obviously in contradiction
 to
  Ivanov's observation \cite{iva02} that ``...electromagnetic mass
 models
  all seem to have negative pressure''. This particular aspect has been
  pointed out in our previous work \cite{ray04a} where we have
 mentioned
  that there are some examples of electromagnetic mass models where
 positive
  pressures are also available, to be shown elsewhere. However, a
  few more examples in this direction need a further study.

\subsection{The Tikekar models}

  We are now interested in discussing the Tikekar \cite{tik84}
 solutions
  which represent a generalization of those of Pant and Sah
 \cite{pan79}. As
  a particular solution, which describes a physically plausible
 distribution
  of a charged perfect gas, Tikekar \cite{tik84} has obtained the
 matter
  density and fluid pressure in the following forms:
\bear
	\rho \eql \frac{b}{4\pi[(b + d +1)^2 - 4d]r^2},
\\
	p \eql \frac{d}{4\pi[(b + d +1)^2 - 4d]r^2},
\ear
  and hence the functional condition is given by
\bear
	\rho + p = \frac{b + d}{4\pi[(b + d +1)^2 - 4d]r^2},
\ear
  where $b$ and $d$ are non-negative constants. These solutions
 \cite{tik84},
  as those of Pant and Sah \cite{pan79}, also suffer from the
 singularity
  problem since the matter density and the fluid pressure are infinite
 at
  the centre $r=0$. This point for both solutions will be discussed
 later on.

  The total mass and charge contained within the sphere of radius
  $a$ are, respectively,
\bear
	m \eql \frac{1}{2}\left[\frac{(b + d)^2 + 2(b - d)}{(b + d +1)^2 -
		4d}\right] a,
\\
	q \eql \left[\frac{(b + d)^2 - 2d}{(b + d +1)^2 - 4d}\right]^{1/2} a.
\ear
  Now, vanishing of the charge in \eq (34) implies that
\bear
	b + d = \pm   \sqrt {2d}.
\ear
  Again, by the use of the pure charge condition, i.e., the
 vacuum-fluid
  equation of state $\rho + p = 0$ \cite{gau85} in \eq (32), one
 obtains
\bear
	b + d = 0.
\ear
  Thus \eqs (35) and (36) imply both $b = 0$ and $d = 0$. Therefore,
  substitutions of these values to \eqs (30), (31) and (33) make all
 the
  physical quantities, including the gravitational mass, vanish and
 provide
  an electromagnetic mass model with the imperfect-fluid equation of
 state
  (i.e., the reduced \eq (32) which now takes the form $\rho + p =0$).

  However, when the constants $b$ and $d$ are identified as
\bear
	b \eql [1 - c(n - 1)^2]/4c,
\\
	d \eql [c(n + 1)^2-1]/4c,
\ear
  the solutions assumes the Pant and Sah form \cite{pan79} which also
  provides an electromagnetic mass model, as we have verified in \sect
 3.1.

\section{Conclusions}

  In the present article, we have shown that some of the known
 solutions,
  related to static charged fluid spheres of Tolman-VI type, are of
 purely
  electromagnetic origin. In this regard, it is also to be mentioned
 that
  Ivanov's solutions \cite{iva02} with a ($\lambda, n$) classification
  scheme are of this category. He has obtained a set of solutions by a
  Bessel function technique, which recover the Pant and Sah solution
  \cite{pan79}. However, in this set of solutions, the density and
 charge
  function are singular at the centre, as stated earlier. It is also
  mentioned that the Tikekar solutions \cite{tik84} suffer from the
 same
  singularity problem. One of the most undesirable features of general
  relativity is the occurrence of spacetime singularities where the
 laws of
  physics break down, and this is inevitable according to the
  Hawking-Penrose singularity theorems \cite{haw73}. However, following
  Herrera and Ponce de Le{\'o}n \cite{her85}, one can consider a sphere
 as
  being composed of (1) a central core of radius $r_0$ inside which all
  physical quantities are finite, and (2) above the core, a
 self-similar
  fluid described by any of the solutions of Sections 3.1 and 3.2. Then
  these solutions can be matched with any suitable interior
 Schwarzschild
  solutions across the surface $r=r_0$ where $0 < r_0 < a$. It is
  interesting to note that, in the context of cosmology, Trautman
  \cite{tra73} suggested that in the Einstein-Cartan theory, where spin
 and
  torsion are taken into account inherently, a singularity is averted
 due
  to the action of torsion in a universe filled with spinning dust.

  Regarding the stability of charged fluid spheres, Bonnor \cite{bon65}
 and
  also independently De and Raychaudhuri \cite{de68} showed that a dust
  cloud of arbitrarily large mass and small radius can remain in
 equilibrium
  if it satisfies the relation between the electric charge and matter
 energy
  densities by $\sigma = \pm  \rho$. However, Glazer \cite{gla76},
  considering radial pulsations, showed that the Bonnor \cite{bon65}
 model is
  electrically unstable. He \cite{gla79} also explicitly established
 the
  effects of electric charge upon dynamical stability. According to
 Stettner
  \cite{ste73}, a fluid sphere of uniform density with a net surface
 charge
  is more stable than a neutral one. Whitman and Burch \cite{whi81},
 for an
  arbitrary charge and mass distribution, showed that a charged
 analogue
  gives more stability. However, application of the pulsation equations
 to
  the charged Pant and Sah solution \cite{pan79} gave an unsatisfactory
  result in connection with the boundary condition which is
 incompatible
  with the densities and pressures \cite{whi81} (a general technique
 for
  studying the stability is provided in the Appendix.

  We would like to point out that the solutions expressed by \eqs (20),
 (22)
  and (24) (except \eq (26) for $n=2$ of Pant and Sah \cite{pan79}), do
 not
  obey the energy condition \cite{haw73}. The reason behind this is
 related
  to the equation of state which, in the present case, takes the form
 $\rho
  + p = 0$ due to application of the {\it ansatz\/} $q(a)/a =
 1/\sqrt{2}$.
  This equation provides two situations:  either $\rho > 0$ and hence
 $p$ is
  negative or $p> 0$ which means that $\rho$ is negative. Thus, the
 strong
  energy conditions $Rij K^i K_j \geq 0$, where $K^i$ is a timelike
 vector,
  are violated in both cases. The first possibility mentioned above
 requires
  that the system should be under tension and hence gravitationally
  repulsive in nature \cite{tiw84}. This case can be expressed in the
 form
  $g_{00}g_{11} = - 1$ and hence $\lambda = -\nu$, which is also
 equivalent
  to the charged de Sitter solution. The second option, viz., $\rho <
 0$
  indicates a negative mass for the inside of the fluid sphere. This is
  possible in the case of Lorentz's extended electron of the size $\sim
  10^{-16}$ cm where the spherically symmetric charged distribution of
 matter
  must contain some negative mass density
 \cite{coo89,bon89,her94,ray04b}.

\section*{Appendix: Stability analysis}
\def\theequation{A.\arabic{equation}}
\sequ 0

  To study the stability, one needs to match the interior solution to
  the exterior Reissner-Nordstr\"{o}m black hole solution
\bearr
	ds^2= \biggl(1 - \frac{2m}{r} + \frac{q^2}{r^2}\biggr) dt^2 -
		\biggl(1 - \frac{2m}{r} + \frac{q}{r^2}\biggr)^{-1}dr^2
\nnn  \inch\cm
 		- r^2 d\Omega_2^2
\ear
  at the junction interface S situated outside the event horizon,  $a >
 r_h
  = m \pm \sqrt{m^2 - q^2 }$. To analyze the stability, it is required
 to
  use the extrinsic curvature, or second fundamental forms, associated
 with
  the two sides of the shell S as $ K_{ij}^\pm =  - u^\pm_{\mu;\nu}
 e^\mu
  _{(i)} e^\nu _{(j)}$, where $u^\pm$ are the unit normals to S and
  $e^\mu _{(i)}$ are the components of the holonomic basis vectors
 tangent
  to S.  According to the Darmois-Israel formalism \cite{isr66,isr67},
 one
  can write the Lanczos equations for the surface stress-energy tensors
  $S_j^i$ at the junction interface $S$ as
  \cite{lob05,lob06,rah06,rah07a,rah07b}
\bear
	S_j^i = - \frac{1}{8\pi} ([K_j^i]  - \delta_j^i K ),
\ear
  where $S_j^i = \diag ( - \varrho , p_{\theta}, p_{\phi}) $ is the
 surface
  energy tensor with $\varrho$ the surface density and $p_\theta$ and
  $p_\phi$ the surface pressures; $[ K_{ij} ] = K_{ij}^+ - K_{ij}^-$
 and
  $ K = [K_i^i ] = {\rm trace}\, [K_{ij}]$.

  To analyze the dynamics of the junction shell, we permit the junction
  radius to become a function of proper time, $a \to a(\tau)$. Now,
 taking
  into account \eq (A.2), one can find
\bear
	\varrho \eql - \frac{1}{4\pi a}
	\biggl[ \sqrt{1 - \frac{2m}{a} + \frac{q^2}{a^2} + \dot{a}^2}-
	\sqrt{c + \dot{a}^2}\biggr],
\\
 	p_{\theta} \eql p_{\phi} =   p =  \frac{1}{8\pi a}
 		\frac{1 - m/a + a \ddot{a} +
          	\dot{a}^2}{\sqrt{1 - 2m/a + q^2/a^2 +\dot{a}^2}}
\nnn\cm
        - \frac{(1 +n)(c+ \dot{a}^2) + a \ddot{a} } {\sqrt{c +
 \dot{a}^2}}
\ear
  In what follows, the overdot means a derivatives with respect to
 $\tau$.

  The  conservation identity $ S^i_{j|i} = - [  \dot{\varrho} +
  2\frac{\dot{a}}{a}( p + \varrho)] $ yields the following relation:
\bear
	\varrho' = -  \frac{2}{a}( p + \varrho)+ Y,
\ear
  where
\bear
	Y =  - \frac{n}{4\pi a^2}  \sqrt{c +\dot{a}^2}.
\ear

  After a little bit of algebra, \eq (A.3) gives the thin shell
  equation of motion
\bear
	\dot{a}^2 + V(a)= 0,
\ear
  where $V(a)$ is the  potential and defined as
\bear
       V(a) =  \frac{1}{2} (f_1 + c)  - 4\pi^2a^2
            \varrho^2 - \frac{(f_1 - c)^2}{64\pi^2a^2 \varrho^2},
\ear
   where $f_1 =   1 - {2m}/{a} + {q^2}/{a^2}$.

  Linearizing around a static solution at $a=a_0$, one can
  expand $V(a)$ around $a_0$ to obtain
\bearr
     V =  V(a_0) + V'(a_0) ( a - a_0)
\nnn  \cm
	+ \frac{1}{2} V''(a_0) ( a - a_0)^2 + O[( a - a_0)^3],
\ear
  where the prime denotes a derivative with respect to $a$.

  Since we are considering linearization around a static solution at $
 a =
  a_0 $, we have $ V(a_0) = 0 $ and $ V'(a_0)= 0 $. Stable equilibrium
  configurations correspond to the condition $ V''(a_0)> 0 $, i.e.,  $
 V(a)$
  has a local minimum at $a_0 $. Now, we define a parameter $\beta$, to
 be
  interpreted as the speed of sound \cite{poi95}, by the relation
\bear
       \beta^2(\varrho) = \frac{ \d p}{\d \varrho}\Big|_\varrho.
\ear
  Following \eq (A.5), we get
\bear
       \beta^2(\varrho) = -1 + \frac{a}{2\varrho'}
       	\biggl[ \frac{2}{a^2} ( p + \varrho) + Y'-\varrho''\biggr].
\ear

\enlargethispage{-8mm}

  The solution is stable if
\bear
    \beta_0^2 < \frac{ A - B + C - S - T  + G - H}{N - L} - 1
\ear
  where $\beta_0 = \beta(a=a_0)$ and  $A$, $B$, $C$, $S$, $T$, $G$,
 $H$,
  $N$, $L$ are given at $a = a_0$ as follows:
\bear
   A \eql  \frac{1}{2}f_1''
     	=  \frac{1}{2}\biggl[\frac{6q^2}{a^4}-\frac{4m}{a^3}\biggr],
\nnv
   B \eql 8\pi^2(\varrho^2-4a\varrho\varrho'-2a^2\varrho'{}^2)
\nn
     \eql \frac{1}{2a^2}(\sqrt{f_1}- \sqrt{c})^2
\nnn \cm
      +\frac{2}{a^2}(\sqrt{f_1} - \sqrt{c})
      	\left(\frac{f_2} {\sqrt{f_1}} -\sqrt{c}\right)
\nnn  \inch
      - \frac{1}{a^2}\biggl(\frac{f_2}{\sqrt{f_1}} - \sqrt{c}\biggr)^2,
\nnv
     C\eql \frac{2}{a^2}[(p+\varrho)+Y']
	\biggl[\frac{(f_1-c)^2}{32\pi^2\varrho^3a^2}-8\pi^2\varrho a^2\biggr]
\nn
        \eql \frac{1}{8\pi a}
		\left(\sqrt{c}(1-n) -\frac{f_2}{\sqrt{f_1}}\right)
\nnn \ \ \
	+ \frac{n\sqrt{c}}{4\pi a^3}
	\biggl[\frac{4\pi}{a}\biggl\{(\sqrt{f_1}- \sqrt{c})-\frac{(f_1 - c)^2}
			{(\sqrt{f_1}- \sqrt{c})^3}\biggr\}\biggr],
\nnv
      S \eql \frac{(f_1')^2}{32\pi^2\varrho^2 a^2}
      	  =\frac{1}{2}\frac{(2m/a^2 - 2q^2/a^3)^2} {(\sqrt{f_1}-
 \sqrt{c})^2},
\nnv
      T \eql \frac{(f_1-c)f_1''}{32\pi^2\varrho^2 a^2}
	=  \frac{1}{2}\frac{(f_1-c)
		(-4m/a^3 + 6q^2/a^4)} {(\sqrt{f_1}- \sqrt{c})^2},
\nnv
      G \eql  \frac{(f_1-c)f_1'}{16\pi^2\varrho^2 a^2}
	\biggl[ \frac{\varrho'}{\varrho}+\frac{\varrho'}{a}+\frac{2}{a}\biggr]
\nn
	\eql \frac{(f_1-c)
	 (2m/a^2 - 2q^2/a^3)}  {(\sqrt{f_1}- \sqrt{c})^2}
\nnn \cm
	\times  \frac{1}{a} \frac{1}{(\sqrt{f_1}- \sqrt{c})}
			\biggl[\biggl(\frac{f_2}{\sqrt{f_1}} -\sqrt{c}\biggr)
\nnn \cm\cm
    	+ \frac{1}{4\pi a^3}\biggl(\frac{f_2}{\sqrt{f_1}}
    		-  \sqrt{c} + \frac{2}{a}\biggr)\biggr],
\nnv
	H \eql \frac{(f_1-c)^2}{16\pi^2\varrho^2 a^2} \biggl[
		\frac{2\varrho'}{a\varrho}+\frac{{3\varrho'}^2}{2\varrho^2}
		+\frac{3}{2a^2}\biggr]
\nn
	 \eql  \frac{(f_1-c)^2} {(\sqrt{f_1}-\sqrt{c})^2}
\nnn\ \ \
	\times\biggl[ \frac{3}{2a^2}- \frac{2}{a^2}
		\frac{1}{(\sqrt{f_1}-\sqrt{c})}
			\biggl(\frac{f_2}{\sqrt{f_1}}-\sqrt{c}\biggr)
\nnn \cm\
	+ \frac{3}{2a^2}\frac{1}{(\sqrt{f_1}- 	\sqrt{c})}
		\biggl(\frac{f_2}{\sqrt{f_1}}-\sqrt{c}\biggr)^2\biggr],
\nnv
      N \eql \frac{4\varrho'}{a^3}\frac{(f_1-c)^2}{32\pi^2\varrho^3
 a^2}
\nn
	\eql \frac{1}{2\pi a^5}
         	\biggl[\frac{f_2}{\sqrt{f_1}} - \sqrt{c}\biggr]
	      \biggl[\frac{(f_1 - c)^2 } {(\sqrt{f_1}- \sqrt{c})^3}\biggr],
\nnv
	L \eql  \frac{32\pi^2\varrho\varrho'}{a }
\nn
	  \eql  -\frac{2}{a^4}(\sqrt{f_1}- \sqrt{c})
			\biggl[\frac{f_2}{\sqrt{f_1}}-\sqrt{c}\biggr],
\ear
  where (see also (A.8))
\bear
  	f_1 \eql 1- 2m/a + q^2/a^2,
\nn
  	f_2 \eql 1- 3m/a + 2q^2/a^2.
\earn

  Thus if $a_0$, $u$, $c$, $m$ and $q$ are specified quantities, then
 the
  stability of the configuration requires the above restriction on the
  parameter $\beta_0$. This means that there exists some part of the
  parameter space where the junction location is stable.

\Acknow
  {One of us (SR) is thankful to the authority of Inter-University
 Centre
  for Astronomy and Astrophysics, Pune, India for providing
 Associateship
  programmes under which a part of this work was carried out. Also UGC
 grant
  (No.  F-PSW-002/04-05/ERO) is gratefully acknowledged. Special thanks
 are
  due to the referees and Dr. F. Rahaman, Jadavpur University, for
 valuable
  suggestions which have enabled us to improve the manuscript
 substantially.}

\end{document}